\documentclass[prl,twocolumn,altaffilletter,numerical,flushbottom,secnumarabic,superscriptaddress,floatfix]{revtex4-2}
\usepackage[utf8]{inputenc}
\usepackage[american]{babel}
\usepackage[T1]{fontenc}
\usepackage[pdftex]{graphicx}  
\usepackage{xcolor}
\usepackage{dcolumn}
\usepackage{comment}
\usepackage{bm}
\usepackage{dsfont}
\usepackage{amsthm}
\usepackage{mathrsfs}
\usepackage{bbold}
\usepackage{braket}
\usepackage{wrapfig}
\usepackage{mathtools}
\usepackage{float}
\usepackage{orcidlink}
\usepackage{physics}
\usepackage[normalem]{ulem}
\usepackage{etoolbox}

\usepackage{hyperref}
\hypersetup{colorlinks,bookmarksopen,bookmarksnumbered,citecolor=red,linkcolor=red,urlcolor=red}

\newcommand{\Id}{\mathbb{1}}
\newcommand{\PauliX}{{\text{X}}}
\newcommand{\PauliY}{{\text{Y}}}
\newcommand{\PauliZ}{{\text{Z}}}

\newcommand{\maj}{\gamma}

\newcommand{\gauss}{\mathcal{G}}
\newcommand{\cliff}{\mathcal{C}}
\newcommand{\cliffgauss}{\mathcal{CG}}

\newcommand{\daga}{^\dag}

\begin{document}

\title{Emergence of Generic Entanglement Structure in Doped Matchgate Circuits}

\author{Alessio Paviglianiti~\orcidlink{0000-0002-2719-7080}}
\affiliation{International School for Advanced Studies (SISSA), via Bonomea 265, 34136 Trieste, Italy}

\author{Luca Lumia~\orcidlink{0000-0002-7783-9184}}
\affiliation{International School for Advanced Studies (SISSA), via Bonomea 265, 34136 Trieste, Italy}

\author{Emanuele Tirrito}
\affiliation{The Abdus Salam International Centre for Theoretical Physics (ICTP), Strada Costiera 11, 34151 Trieste, Italy}
\affiliation{Dipartimento di Fisica ``E. Pancini", Universit\`a di Napoli ``Federico II'', Monte S. Angelo, 80126 Napoli, Italy}

\author{Alessandro Silva}
\affiliation{International School for Advanced Studies (SISSA), via Bonomea 265, 34136 Trieste, Italy}

\author{Mario Collura~\orcidlink{0000-0003-2615-8140}}
\affiliation{International School for Advanced Studies (SISSA), via Bonomea 265, 34136 Trieste, Italy}
\affiliation{INFN Sezione di Trieste, 34136 Trieste, Italy}

\author{Xhek Turkeshi~\orcidlink{0000-0003-1093-3771}}
\affiliation{Institute f\"{u}r Theoretische Physik, Universit\"{a}t zu K\"{o}ln, Z\"{u}lpicher Straße 77, D-50937 K\"{o}ln, Germany}

\author{Guglielmo Lami~\orcidlink{0000-0002-1778-7263}}
\affiliation{Laboratoire de Physique Th\'eorique et Mod\'elisation, CNRS UMR 8089, CY Cergy Paris Universit\'e, 95302 Cergy-Pontoise Cedex, France}

\begin{abstract}
Free fermionic Gaussian, a.k.a. matchgate, random circuits exhibit atypical behavior compared to generic interacting systems. They produce anomalously slow entanglement growth, characterized by diffusive scaling $S(t) \sim \sqrt{t}$, and evolve into volume-law entangled states at late times, $S \sim N$, which are highly unstable to measurements. Here, we investigate how doping such circuits with non-Gaussian resources (gates) restores entanglement structures of typical dynamics. We demonstrate that ballistic entanglement growth $S(t) \sim t$ is recovered after injecting an extensive total amount of non-Gaussian gates, also restoring Kardar-Parisi-Zhang fluctuations. When the evolution is perturbed with measurements, we uncover a measurement-induced phase transition between an area-law and a power-law entangled phase, $S \sim N^\alpha$, with $\alpha$ controlled by the doping. A genuine volume-law entangled phase is recovered only when non-Gaussian gates are injected at an extensive rate. Our findings bridge the dynamics of free and interacting fermionic systems, identifying non-Gaussianity as a key resource driving the emergence of non-integrable behavior.
\end{abstract}

\maketitle

\textit{Introduction --}
Random quantum circuits offer a powerful framework for exploring many-body dynamics and the transition from classically simulable to genuinely complex quantum evolution~\cite{fisher2023random}. This approach has been central in revealing key dynamical phenomena, including chaos, thermalization, and entanglement spreading, in both unitary and monitored settings~\cite{potter2022entanglement,li2025measurement,lunt2022quantum}. 
In generic random unitary circuits, entanglement grows linearly in time, exhibiting fluctuations governed by the Kardar-Parisi-Zhang (KPZ) universality class~\cite{Nahum_2017,Zhou_2019,zhou2020entanglement,sierant2023entanglement,sommers2024zero,ha2025rougheningtransitionquantumcircuits}, and reaches a volume-law scaling in the steady state. When the dynamics is perturbed by measurements, this behavior persists under weak monitoring, but gives way to an area-law entangled phase beyond a critical measurement rate, signaling the onset of a measurement-induced phase transition (MIPT)~\cite{Li_2019,Skinner_2019,ippoliti2020mipt,choi2020mipt,gullans2020mipt,gullans2020scalable,jian2020mipt,baobao,zabalo2020critical,turkeshi2022measurement,sierant2022universal,sierant2022measurement,Sierant_2023,Lami_2024_1,zabalo2022operator,ippoliti2024learnability,agrawal2024observing}.

Restricting the set of quantum gates, either due to hardware limitations or resource-theoretic constraints, can alter quantum dynamics dramatically, potentially enabling efficient classical simulability. A relevant example is fermionic Gaussian circuits, or matchgates~\cite{Valiant_2001,Terhal_2002,Jozsa_2008,Mocherla_2024}, which describe the dynamics of non-interacting fermions. These arise naturally in condensed matter~\cite{Fradkin_2013}, quantum chemistry~\cite{Szabo_1996,Mcardle_2020}, and fermionic linear optics~\cite{Bravyi_2004,Knill_2011,Oszmaniec_2022,Dias_2024,Reardonsmith_2024}. Random matchgate circuits can be simulated efficiently via two-point correlations~\cite{Surace_2022,Mbeng_2024}, and exhibit diffusive entanglement growth saturating to a volume-law scaling~\cite{Nahum_2017,roosz,swann2023spacetimepictureentanglementgeneration,nahum2020entanglement}. In contrast to generic circuits, this entanglement phase is extremely fragile: any finite measurement rate collapses it to a logarithmic or area-law regime~\cite{Cao_2019,Alberton_2021,Turkeshi_2021,Turkeshi_2022,Muller_2022,Paviglianiti_2023,bucchold2022mipt,Coppola_2022,Fava_2023,PhysRevB.108.L020306,Tiutiakina2025fieldtheory,chahine2024entanglementphaseslocalizationmultifractality,poboiko2025mit,poboiko2023theory,poboiko2024mipt,xing2024}, indicating that free fermionic systems are unable to store quantum information robustly~\cite{Fidkowski2021howdynamicalquantum}.

\begin{figure}[t!]
\includegraphics[width=\columnwidth]{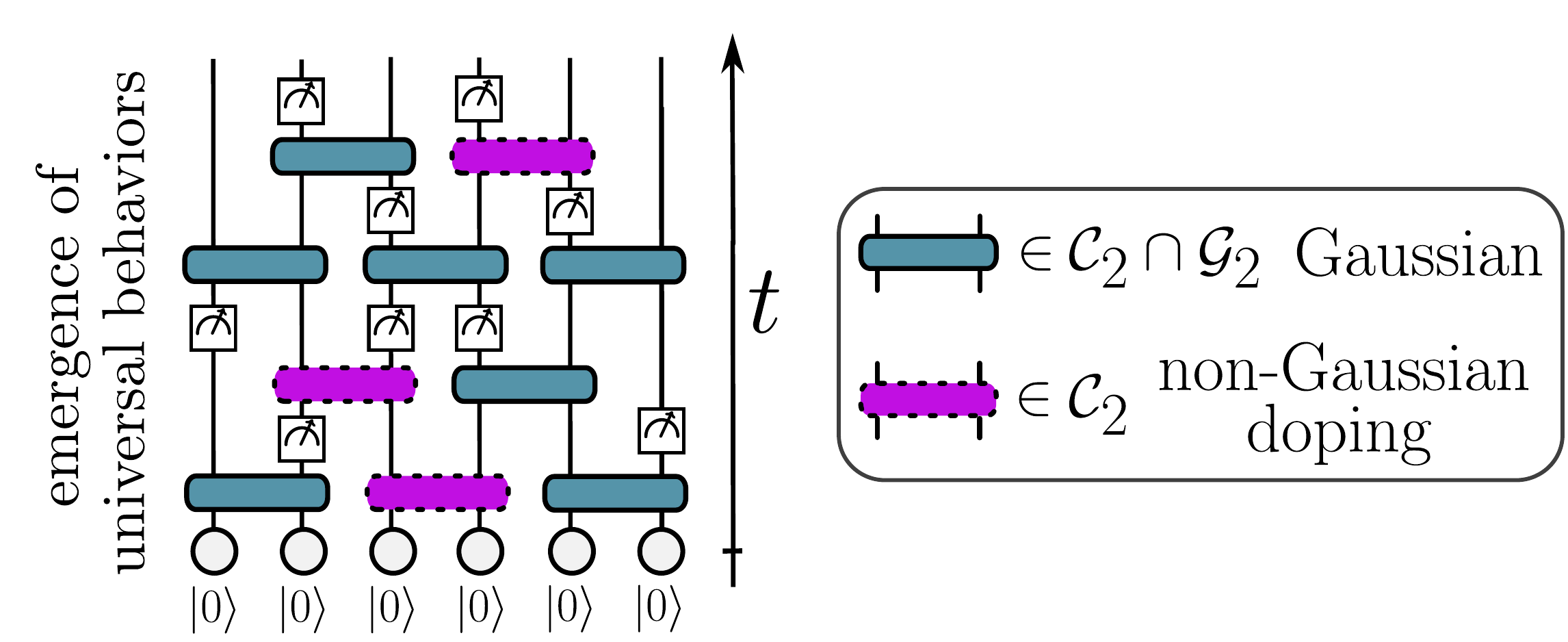}
\caption{\label{fig:cartoon} Schematic of the monitored circuits studied in this work, alternating layers of two-qubit gates and single-qubit projective measurements. Gates are either Gaussian (teal) or non-Gaussian (dotted purple), both drawn from the Clifford group. By tuning the density of non-Gaussian gates, we probe the onset of universal behavior beyond the free-fermion limit.}
\end{figure}

\begin{figure*}[t!]
\includegraphics[width=0.95\textwidth]{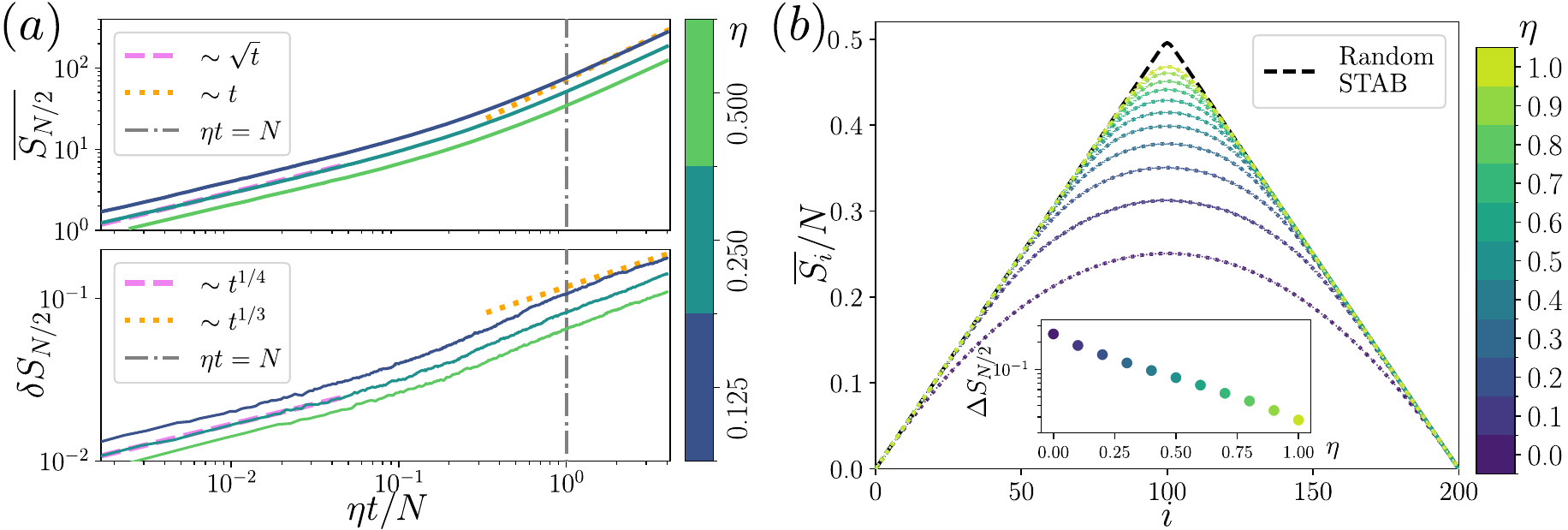}
\caption{\label{fig:plot_3} 
$(a)$ Evolution of the half-chain entanglement entropy $\overline{S_{N/2}}$ (top) and its fluctuations $\delta S_{N/2}$ (bottom). We consider unitary dynamics ($p=0$) and different rates of non-Gaussianity injection by varying $\eta$ while fixing $\beta=1$. Data for $N=600$ averaged over $3000$ random realizations. At late times, the injection of non-Gaussianity restores linear growth $\overline{S_{N/2}(t)}\sim t$ and KPZ-like fluctuations $\delta S_{N/2}\sim t^{1/3}$ , characteristic of generic random circuits. $(b)$ Page curve for increasing values of non-Gaussian doping, showing the entropy density $\overline{S_i}/N$ as a function of the subsystem length $i$. Data for $N=200$ and $N_\mathrm{NG}=5N\eta$ injected non-Gaussian gates, averaged over $3000$ random realizations. The dashed black line represents the Page curve $\overline{S_{i}}\rvert_{\cliff}$ of random stabilizer states. Inset: deviation $\Delta S_{N/2} = \overline{S_{N/2}}\rvert_{\cliff} - \overline{S_{N/2}}$ as a function of $\eta$.}
\end{figure*}

Matchgate circuits are unable to induce fermionic interactions and cannot generate universal quantum complexity~\cite{Marian_2013, Lami_2018, Hebenstreit_2019, Lumia_2024, Lyu_2024,morralyepes2025disentanglingstrategiesentanglementtransitions, Sierant_2025,mele2025efficient}. Accessing generic many-body behavior requires the injection of non-Gaussian resources~\cite{Chitambar_2019, Lumia_2024} via unitaries outside the matchgate class, analogous to magic-state injection in Clifford (stabilizer) circuits~\cite{zhou2020single,Leone2021quantumchaosis,Oliviero_2021,Magni_2025,beri,tarabunga2024magictransitionmeasurementonlycircuits,liu2024classicalsimulabilitycliffordtcircuits,fux2024entanglement,szombathy2025spectralpropertiesversusmagic,szombathy2025independentstabilizerrenyientropy,Niroula_2024,Turkeshi2024c,trigueros2025nonstabilizernesserrorresiliencenoisy}. While this paradigm is well established for stabilizer circuits, its fermionic counterpart remains largely unexplored. 
Here, we address this problem by studying \textit{doped} matchgate circuits that interpolate between Gaussian and generic features of quantum dynamics. 
We focus on two central questions: \textit{(i) How much non-Gaussianity is required to restore ballistic entanglement growth and KPZ fluctuations?} \textit{(ii) Can non-Gaussian doping stabilize a volume-law entanglement phase in presence of measurements?}
To answer, we investigate hybrid Clifford circuits in which matchgate dynamics are interspersed with a tunable density of non-Gaussian gates, enabling a controlled crossover from integrable to generic behavior. 

In the unitary case, we find that even an intensive injection of non-Gaussian gates per unit time significantly enhances entanglement growth, driving a crossover from diffusive to ballistic scaling and restoring KPZ fluctuations once the total number of non-Gaussian gates becomes extensive in system size. 
In monitored circuits, we instead observe that non-Gaussian doping generally gives rise to a sub-volume (power-law) entangled phase, and an \emph{extensive} injection of non-Gaussianity per unit time is required to stabilize the volume-law regime. Our results quantify the minimal non-Gaussian resources needed to push non-interacting circuits into a genuinely complex regime, establishing the role of integrability breaking in the emergence of generic entanglement structures from free fermionic behavior.

\textit{Setup --} We now introduce the circuit model we study, realized with fermionic Gaussian and non-Gaussian gates. Given a system of $N$ qubits, let $\mathcal{P}_N=\{ \Id, \PauliX, \PauliY, \PauliZ \}^{\otimes N}$ be the Pauli group. Using the Jordan-Wigner mapping, we define the $2N$ Majorana operators $\maj_{2i-1} = \PauliZ_1 \cdots \PauliZ_{i-1} \PauliX_i$, $\maj_{2i} = \PauliZ_1 \cdots \PauliZ_{i-1} \PauliY_i$, satisfying the canonical anticommutation relations $\{ \maj_\mu, \maj_\nu \} = 2 \delta_{\mu\nu}$~\cite{Surace_2022,Mbeng_2024}. Unitaries $U$ that act linearly on Majorana operators such as $U^\dagger \maj_\mu U = \sum_\nu Q_{\mu,\nu}\maj_\nu$, for a certain orthogonal transformation $Q\in \mathrm{SO}(2N)$ , are called \textit{matchgates}, and form the group $\gauss_N$ of Gaussian operations. 
When acting on $\ket{0}^{\otimes N}$, these gates generate the fermionic Gaussian states (FGSs), a special class of quantum states describing non-interacting fermions. Any FGS $|\psi\rangle$ is completely characterized by their Majorana correlation functions $M_{\mu\nu}=-i\langle\psi| [\maj_{\mu},\maj_{\nu}]|\psi\rangle/2$, enabling efficient classical simulation~\cite{Surace_2022,Mbeng_2024}.

Matchgate circuits are not universal for quantum computation, as they lack a crucial quantum resource known as non-Gaussianity (or fermionic magic)~\cite{Sierant_2025}. This may be introduced either through the injection of specially prepared fermionic magic states~\cite{Hebenstreit_2019}, or by enabling operations beyond the matchgate framework, which act non-linearly on Majorana operators and effectively induce fermionic interactions~\cite{Lumia_2024}~\footnote{This closely parallels the role of non-stabilizer “magic states” or gates in promoting Clifford circuits to full quantum universality.}. As a result, non-Gaussian unitaries enable entanglement structures and computational complexity that are fundamentally inaccessible to purely Gaussian dynamics.

For practical convenience, we constrain all gates (whether Gaussian or not) to lie within the Clifford group $\cliff_N$, the set of unitaries that map the Pauli group $\mathcal{P}_N$ onto itself up to phases. 
Clifford circuits generate stabilizer states $\ket{\psi}$, which are uniquely defined by $N$ independent Pauli strings $P_i \in \mathcal{P}_N$ satisfying $P_i \ket{\psi} = \pm \ket{\psi}$. 
This structure enables efficient classical simulation, allowing us to probe both Gaussian and non-Gaussian dynamics for system sizes of several hundred qubits. Hence, all numerical simulations in this Letter are performed using the stabilizer formalism~\cite{hamma2005bipartite,Hamma_2005}.
While the Clifford restriction might appear limiting, it is important to note that many nonlinear properties, such as entanglement dynamics and state-design behavior~\cite{Webb_2016, Zhu_2017,Mele_2024,Wan_2023,Braccia_2025}, closely mirror those of Haar-random circuits~\footnote{The Clifford group $\cliff_N$ forms a unitary 3-design~\cite{Webb_2016, Zhu_2017} for the Haar ensemble~\cite{Mele_2024}, meaning that random unitaries sampled from these sets give rise to equal expectation values of observables acting on at most three replicas of the system. Similarly, the set \emph{Clifford-Gaussian} forms a $3$-design for $\gauss_N$~\cite{Wan_2023, Braccia_2025}.}. Clifford circuits therefore provide a controlled framework to approximate the entanglement and dynamical properties of Gaussian and non-Gaussian states.
Importantly for what follows, we note that any Clifford-Gaussian transformation $U \in \cliffgauss_N = \cliff_N\cap\gauss_N$ acts on Majorana operators as a signed permutation $U\daga\maj_\mu U = \pm \maj_{\sigma(\mu)}$, where $\sigma\in S_{2N}$ is a permutation of $2N$ elements~\footnote{Among the $|\cliff_2|=11520$ gates, Clifford matchgates are only $|\cliffgauss_2|=192$.}.

Fig.~\ref{fig:cartoon} illustrates our setup: a system initialized in the vacuum $\ket{\psi_0} = \ket{0}^{\otimes N}$ evolves under a brickwall circuit composed of alternating layers of two-qubit gates and measurements. 
In the latter, every qubit is measured in the computational basis with probability $p$, and left untouched with probability $1-p$. 
In the unitary layer, each gate $U$ is drawn uniformly from $\cliff_2$ with probability $q$, and from the Gaussian matchgates $\cliffgauss_2$ with probability $1-q$. 
The resulting number of non-Gaussian gates per layer scales as $O(qN)$. By setting $q = \eta / N^\beta$, we enable control over this scaling: the number of non-Gaussian gates per layer is extensive for $\beta = 0$, sub-extensive for $0 < \beta < 1$, and intensive for $\beta = 1$.

\emph{Unitary dynamics and Page Curve --} 
We begin by analyzing the unitary dynamics ($p=0$), focusing on the bipartite entanglement entropy $S(\rho_A) = -\mathrm{Tr}(\rho_A \log_2 \rho_A)$, where $\rho_A = \mathrm{Tr}_{\overline{A}}[ |\psi\rangle\langle\psi|]$ is the reduced density matrix for a subsystem $A$ marginalizing its complementary $\overline{A}$. 
We study the half-chain $A=\{1,\dots,N/2\}$ entanglement entropy $S_{N/2}\equiv S(\rho_A)$ as a function of time (i.e., circuit depth) $t$ for intensive non-Gaussian injection ($\beta = 1$).

First, in the purely Gaussian case of $\eta=0$, entanglement shows diffusive growth $S_{N/2} \sim \sqrt{t}$, which is typical of free fermionic systems subject to noise or disorder~\cite{Nahum_2017,PhysRevB.93.134305}. In Fig.~\ref{fig:plot_3}(a), we present our numerical results in presence of non-Gaussian doping. At early times, entanglement preserves the diffusive growth of free fermions, and gradually shifts toward a ballistic scaling typical of generic quantum dynamics. This transition occurs when the total number of non-Gaussian gates, $N_{\mathrm{NG}} \propto q N t = \eta t$, becomes extensive, i.e., $N_{\mathrm{NG}} \sim N$ [see dotted grey line in Fig.~\ref{fig:plot_3}(a)]. 

\begin{figure}[t!]
\includegraphics[width=0.95\columnwidth]{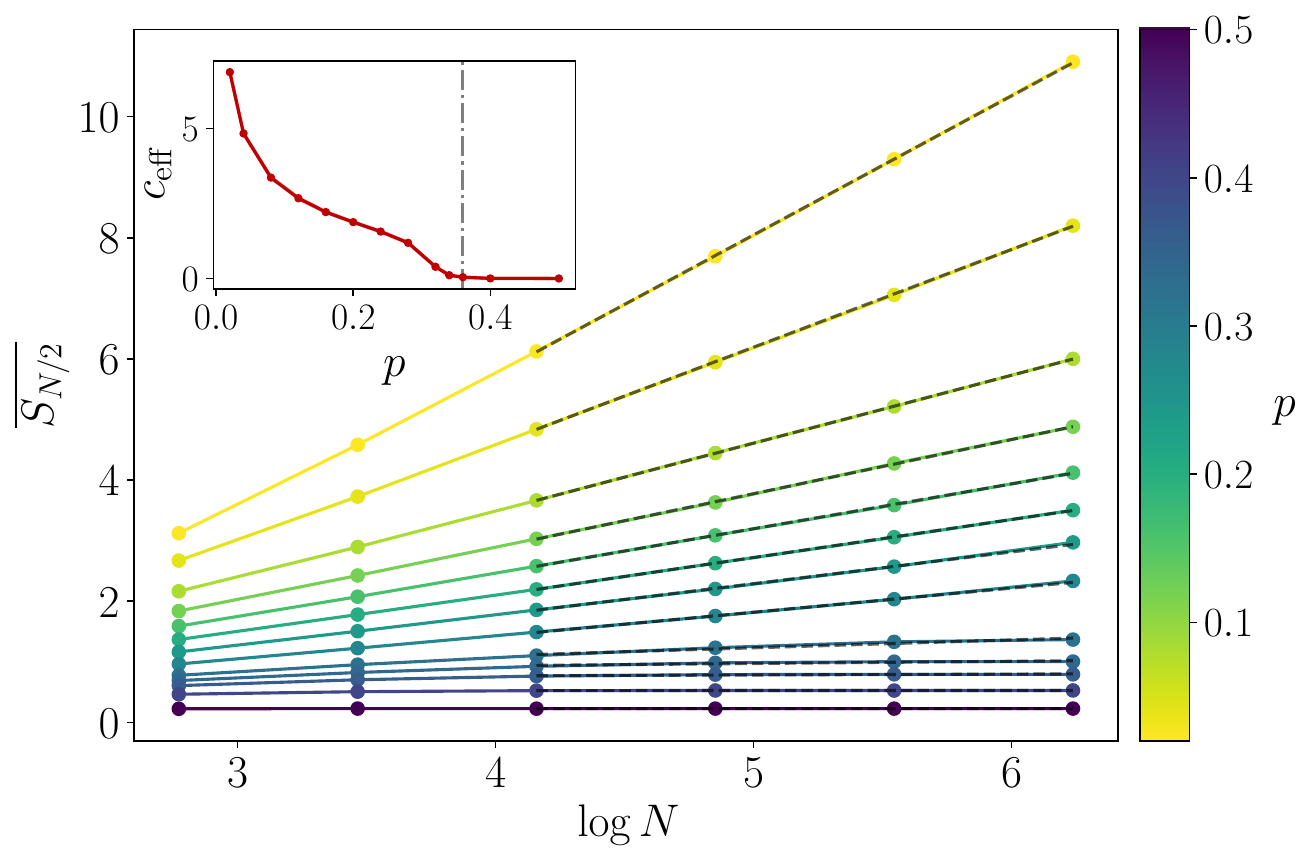}
\caption{\label{fig:gaussian_mipt} Entanglement transition in Gaussian Clifford circuits. For small values of $p$, the system exhibits logarithmic scaling of the long-time entanglement entropy $\overline{S_{N/2}} \approx c_{\mathrm{eff}}/3 \log N$. As $p$ increases above a critical $p_c|_\cliffgauss\approx 0.36$, the entropy attains an area-law, $\overline{S_{N/2}} \sim \mathrm{const}$. Inset: prefactor $c_{\mathrm{eff}}$ of the logarithmic scaling as a function of $p$. All data are averaged over $1000$ random circuit realizations.}
\end{figure}  

The diffusive Gaussian growth can be understood analytically by mapping the purely Gaussian circuit ($\eta=0$) to a classical model of arcs. The initial state $\ket{0}^{\otimes N}$ is stabilized by $P_i = \PauliZ_i = -i \maj_{2i-1} \maj_{2i}$, which we represent as a collection of arcs connecting adjacent “Majorana points” $(2i, 2i+1)$ along a $2N$-site chain. As observed previously, Gaussian matchgates permute Majorana operators as $U^\dagger P_i U = -i \maj_{\sigma(2i)} \maj_{\sigma(2i+1)}$, which amounts to permuting arc endpoints. These rules are illustrated in the following sketch, together with an example showing how a circuit can be unrolled into an arc configuration:
\begin{equation}\label{eq:cartoon_arcs_unitary}
\vcenter{\hbox{\includegraphics[width=0.65\linewidth]{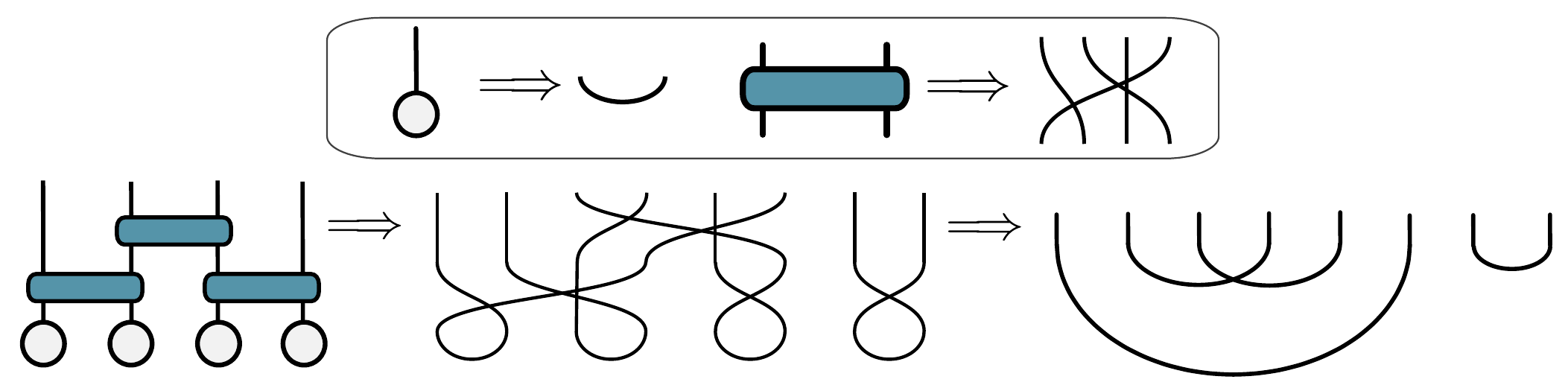}}}
\end{equation}
Crucially, although the arc representation ignores the signs of $P_i$, it fully encodes the entanglement structure: for any bipartition, $S(\rho_A)$ equals half the number of arcs crossing between $A$ and its complement $\overline{A}$. 

Since the matchgates (and thus the corresponding permutations) are picked randomly, each arc endpoint moves stochastically. Let $P_t(x|x_0)$ denote the probability that an arc endpoint initially at $x_0$ reaches $x$ after $t$ steps. As shown in the End Matter, this distribution is binomial, and approaches the normal distribution $P_t(x|x_0) \approx \mathcal{N}(x;x_0,4t)$ in the limit of $N \gg \sqrt{t}\gg 1$. This implies diffusive motion of the endpoints, giving rise to a mean arc size growing as $\sim \sqrt{t}$. Most importantly, $P_t(x|x_0)$ enables the calculation of the average entanglement entropy, yielding the exact result $\overline{S_{N/2}(t)}\approx \sqrt{t/\pi}$, with a universal prefactor that differs from the diffusive pairing dynamics of Majorana defects~\cite{Nahum_2020}. At long times, $P_t(x|x_0)$ approaches the uniform distribution, causing entanglement to saturate to the volume-law $\overline{S_{N/2}} = N/4 + O(1)$.

\begin{figure}[t!]
\includegraphics[width=0.45\textwidth]{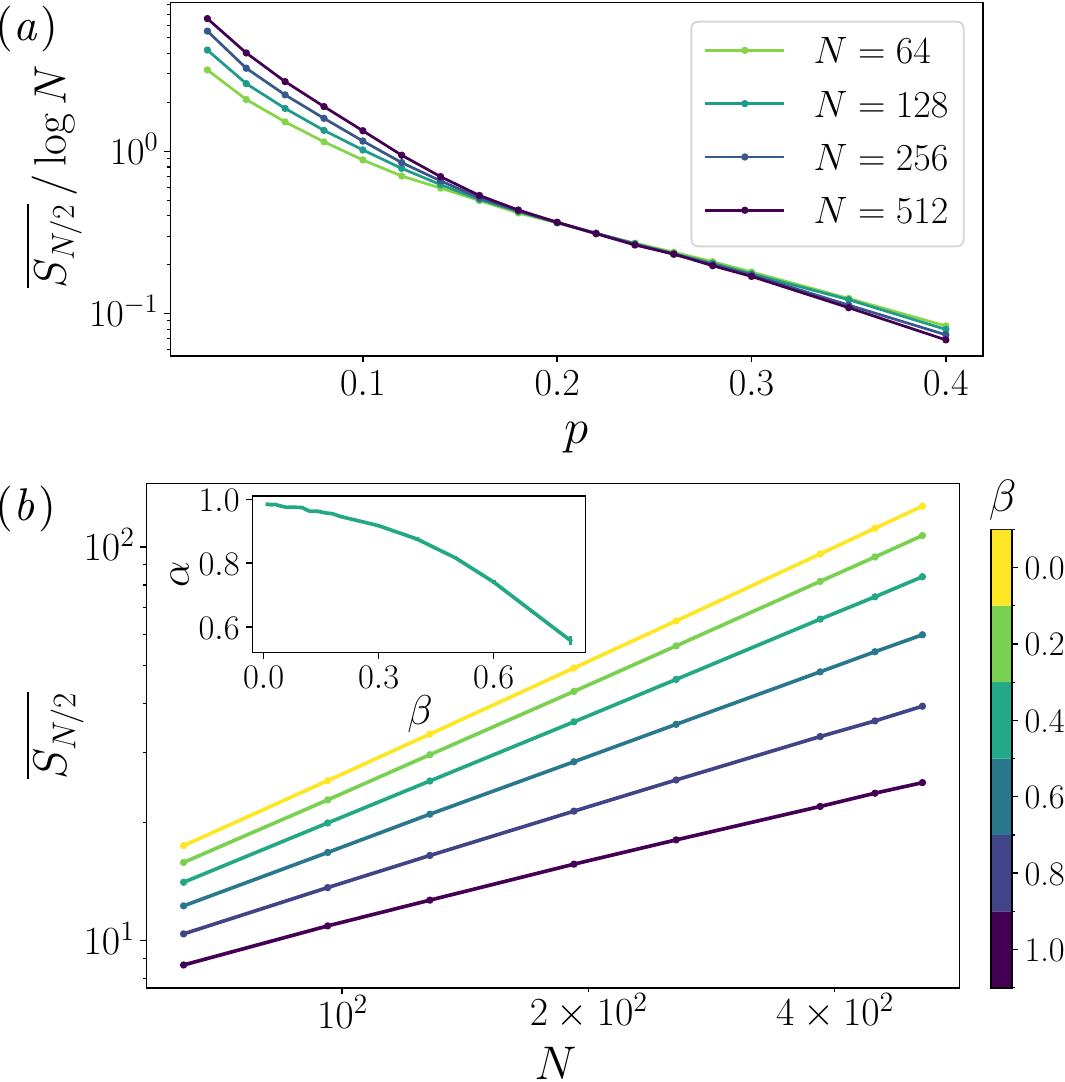}
\caption{
\label{fig:monitored_vertical}
$(a)$ Stationary entanglement for a monitored circuit with sub-extensive doping of non-Gaussian resources at $\eta=1,\,\beta=0.5$, as a function of the measurement rate. The crossing point $p_c\approx0.21$ identifies a MIPT between a region with a super-logarithmic entanglement growth.  $(b)$ Half-chain entanglement entropy for different system sizes $N$ and different doping exponents $\beta$, having fixed $p=0.01,\,\eta=1$. Every $\beta<1$ results in a power-law with exponent $\alpha<1$, as shown explicitly in the inset. Data obtained by averaging over $500$ quantum trajectories.
}
\end{figure}

In Fig.~\ref{fig:plot_3}(a), we also analyze the fluctuations of the entanglement entropy, $\delta S_{N/2}$ across different circuit realizations.  
Initially, they scale as $t^{1/4}$, consistent with the variance of a binomial distribution with mean $\propto \sqrt{t}$. At late times, they shift to the $\sim t^{1/3}$ scaling characteristic of the KPZ universality class~\cite{Nahum_2017,Zhou_2019}, originally discovered in the context of stochastic surface growth~\cite{PhysRevLett.56.889}. 
As before, this crossover occurs once the number of non-Gaussian gates becomes extensive, $N_\mathrm{NG} \sim N$.

The long-time entanglement entropy achieved in unitary random circuits is known to follow the universal Page curve~\cite{Page_1993,Foong_1994,Ruiz_1995,Sen_1996}, which is different for Clifford~\cite{Dahlsten_2007} and Clifford Gaussian gate classes. We investigate the crossover between the two induced by non-Gaussian doping. Only for this study, we initialize the system in a random Gaussian stabilizer state (corresponding to a random arc configuration), and we evolve it with the doped unitary circuit for a time $T=5N$ using variable rate $\eta$ and fixing $\beta=1$. This yields a total non-Gaussian content $N_\mathrm{NG}/N = 5\eta$. In Fig.~\ref{fig:plot_3}(b) we show the average entanglement density $\overline{S}_i/N$ of a compact subsystem of length $i\in[1,N]$. As the doping increases, the Page curve converges smoothly from the Clifford Gaussian page curve (derived from the arc model) to the stabilizer one (known analytically). The deviation at half-chain, $\Delta S_{N/2}=\overline{S_{N/2}}\rvert_{\cliff} - \overline{S_{N/2}}$, decays approximately exponentially as $\Delta S_{N/2} \sim e^{-a N_\mathrm{NG}/N}$ with $a > 0$.

Our results indicate that injecting $N_\mathrm{NG} = O(N)$ non-Gaussian gates suffices to restore typical entanglement features. However, as we will see, this no longer holds true once measurements are introduced ($p > 0$).

\emph{Gaussian monitored dynamics --} We now consider fully Gaussian monitored Clifford circuits without any doping ($\eta=0$). Unlike the unitary case, where entanglement grows extensively, any nonzero measurement rate $p > 0$ destroys the volume-law scaling. In detail, under weak monitoring the long-time entanglement entropy scales logarithmically with system size, $\overline{S_{N/2}} \sim c_{\mathrm{eff}}/3 \log N$, and undergoes a transition to an area-law phase $\overline{S_{N/2}} \sim \mathrm{const}$ above a critical point $p_c|_\cliffgauss\approx 0.36$. This behavior, shown in Fig.~\ref{fig:gaussian_mipt}, stands in contrast with that of monitored Clifford circuits, which display a stable volume-law phase for $p<p_c|_\cliff \approx 0.16$~\cite{Li_2019,sierant2022measurement}.

The emergent logarithmic phase can be explained using the arc model. First, consider the action of a measurement in the computational basis on the stabilizers $P_i$. If $\ket{\psi}$ is stabilized by $i\maj_{2i-1}\maj_{j}$ and $i\maj_{2i}\maj_{k}$, then a measurement of $\PauliZ_i$ replaces these with $i\maj_j\maj_k$ and $i\maj_{2i-1}\maj_{2i}$, regardless of the outcome.
This operation “glues” the endpoints $2i-1$ and $2i$ of the original arcs, forming a new arc $(j,k)$, as well as a local arc $(2i-1,2i)$. Graphically, this mechanism can be represented by the diagram $\raisebox{-0.5em}{\includegraphics[width=0.15\linewidth]{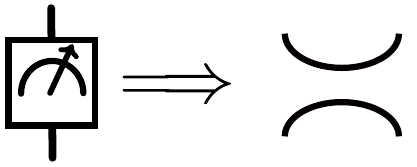}}$, providing an additional rule to update arc configurations under measurement. 

The long-time entanglement scaling in presence of weak monitoring $p\ll 1$ can be captured by an effective master equation for the arc-length distribution $P_t(\ell)$, where $\ell = x - y$ denotes the length of an arc measured in physical sites at time $t$ (see End Matter for details on the derivation and solution). The circuit of Fig.~\ref{fig:cartoon} alternates between a unitary and a measurement layer. The first one introduces a diffusive term: each arc endpoint $x,y$ either stays the same with probability $1/2$ or shifts by $\pm 1$ with probability $1/4$ each. In contrast, measurements merge two arcs of lengths $\ell$ and $\ell'$ into one localized arc of length $0$ and a composed arc of length $\ell + \ell'$. Since we are tracking a single-arc distribution $P_t(\ell)$, we model this by assuming that a measurement resets $\ell$ to zero with probability $1/2$, whereas it updates it to $\ell + \ell'$ (with $\ell'$ sampled from the distribution itself) with the complementary probability $1/2$.

This description, valid at long times and low measurement rates, yields a non-linear master equation of the form $P_{t+1}(\ell) = \mathcal{K}[P_t](\ell)$, where $\mathcal{K}$ is a suitable kernel function (see End Matter for its explicit form). Remarkably, the steady-state solution $P(\ell)=\lim_{t\to\infty}P_t(\ell)$ can be determined analytically in momentum space after introducting $\Tilde{P}(k) = \sum_\ell e^{-i k \ell} P(\ell)$, where we find
\begin{equation}\label{master_equation_k}
    p \Tilde{P}^2(k) + [1-2p-\cos^{-4}(k/2)]\Tilde{P}(k) + p = 0.
\end{equation}
Solving this equation, the long-length behavior is set by $\Tilde{P}(k)\approx 1-|k|/\sqrt{2p}$ for $|k|\ll 1$, which implies the asymptotic form $P(\ell)\approx 1/(\pi\sqrt{2p}\ell^2)$ for $\ell\to \infty$. Lastly, this arc-length distribution enables the direct evaluation of entanglement, yielding $\overline{S_{N/2}}\sim \log N$. Thus, the arc model provides a clear analytic explanation for the emergent logarithmic phase of monitored free fermions.

We point out that the arc model closely resembles the diffusive dynamics of Majorana defects~\cite{Nahum_2020}, but two-site interactions introduce crossings between Majorana worldlines, as in completely packed loop model with crossings (CPLC)~\cite{Nahum_2013, Sang_2021, Merritt_2023,buchhold2024mipt}. These models are reported to exhibit an additional $\sim \log^2 N$ contribution to the entanglement entropy, which becomes numerically visible only for very large system sizes, $N \gtrsim 10^4$~\cite{Sang_2021}. This term also arises from renormalization-group arguments and field-theoretical calculations~\cite{Fava_2023,Tiutiakina2025fieldtheory}. We believe that a similar deviation could be present in the arc model considered here; however, the master equation we develop does not capture this contribution.

\emph{Doped monitored dynamics --} Finally, we study the full entanglement phase diagram in the presence of both measurements ($p > 0$) and non-Gaussian doping ${q = \eta/N^\beta > 0}$. In Fig.~\ref{fig:monitored_vertical}(a), we show the late-time half-chain entanglement $\overline{S_{N/2}}$ versus $p$ for various system sizes, fixing $\eta=1$ and $\beta=0.5$. The crossing of curves at different $N$ signals the presence of a MIPT. Above the critical point we observe the area-law, instead, for $p<p_c$, interestingly the entanglement is sub-extensive, scaling as a power law $\overline{S_{N/2}}\sim N^\alpha$. This already highlights a key difference from the unitary case, where an intensive doping with $\beta=1$ was sufficient to restore the properties of random Clifford circuits.

To probe the possibility of recovering the volume-law phase (i.e., $\alpha=1$), we explore how steady-state entanglement is influenced by $\beta$. This is shown in Fig.~\ref{fig:monitored_vertical}(b) for fixed $\eta=1$ and $p=0.01$, where we observe that entanglement features the power-law scaling $\sim N^\alpha$ for all choices of $\beta$, with the exponent $\alpha$ depending on $\beta$ (and, in general, on $p$ as well). Importantly, our analysis of the scaling behavior (accounting for logarithmic corrections expected in stabilizer circuits~\cite{Li_2019,sierant2022measurement}) highlights that volume-law entanglement is recovered only for $\beta=0$ (see inset), thus requiring extensive non-Gaussianity injection per unit time to fully stabilize the volume-law phase of Clifford circuits. This is a consequence of the monitoring: measurements can destroy non-Gaussianity by projecting the system into locally Gaussian states, partially restoring Gaussian entanglement properties.

Near the transition, the entanglement obeys the scaling form $\overline{S_{N/2}}(p)=a(p_c)\log N + F[(p-p_c)N^{1/\nu}]$, allowing a collapse onto a universal curve $F$ to extract $p_c$ and the correlation length exponent $\nu$.
At $\eta=0$, the system belongs to the universality class of CPLC~\cite{Nahum_2013, Merritt_2023, Sang_2021, buchhold2024mipt}.
Interestingly, for $\beta=0$ we find that any $\eta>0$ yields the same critical behavior with $\nu \approx 1.3$, thus placing the model in the Clifford MIPT universality class~\cite{sierant2022measurement}.

\emph{Conclusions --} In this work, we have demonstrated how doping matchgate circuits with non-Gaussian resources restores key entanglement features of generic many-body dynamics. Leveraging an exact mapping to a classical arc model, we provide an analytical understanding of the purely Gaussian case, capturing both diffusive entanglement growth and the emergence of logarithmic scaling under measurements. By injecting an extensive number of non-Gaussian gates, we recover ballistic entanglement growth and KPZ-class fluctuations in unitary circuits, along with the characteristic Page curve of generic states.

In the monitored setting, we uncover a measurement-induced phase transition between area-law and power-law entangled phases. Crucially, a genuine volume-law phase emerges only when the non-Gaussian doping is extensive per unit time. Our results quantify the minimal fermionic magic required to drive circuits beyond the non-interacting limit into a genuinely complex regime, establishing a direct connection between quantum complexity, as given by non-Gaussianity, and the structure of entanglement.

A topic of key interest for future research is the quantification of non-Gaussianity as a resource and the study of its phase diagram, which may also feature critical behaviors. In this direction, we ask whether it is possible to develop an analytic framework also in the presence of doping, given that the dynamics remains `simple' in the sense of stabilizers. This is left for follow-up investigations.

\subsection*{Acknowledgments}
The authors acknowledge valuable discussions with Jacopo De Nardis, Gerald Fux, Andrea Gambassi, Piotr Sierant.
E.T. acknowledges support from ERC under grant agreement 
n.101053159 (RAVE), and CINECA (Consorzio Interuniversitario per il Calcolo Automatico) award, under the 
ISCRA initiative and Leonardo early access program, for 
the availability of high-performance computing resources 
and support.
A.S. acknowledges the support of the grant PNRR MUR project PE0000023-NQSTI, and of the project “Superconducting quantum-classical linked computing systems (SuperLink)”, in the frame of QuantERA2 ERANET COFUND in Quantum Technologies.
X.T. acknowledges DFG Collaborative Research Center (CRC) 183 Project No. 277101999 - project B01 and DFG under Germany's Excellence Strategy – Cluster of Excellence Matter and Light for Quantum Computing (ML4Q) EXC 2004/1 – 390534769. 
G.L. was supported by the ERC Starting Grant 101042293 (HEPIQ) and the ANR-22-CPJ1-0021-01.
This work was supported by the PNRR MUR project PE0000023-NQSTI, and the PRIN 2022 (2022R35ZBF) - PE2 - ``ManyQLowD''.

\subsection*{Author Contributions}

A.P., L.L., and G.L. conceived the initial idea and performed the numerical simulations. A.P. and G.L. carried out the analytical calculations. G.L. and X.T.
supervised the research. All authors contributed to
the development of the project, discussed the results, and
contributed to writing and revising the manuscript.

\bibliography{biblio_cliffgauss}

\pagebreak
\widetext
\newpage
\begin{center}
\textbf{\large End Matter}
\end{center}

\setcounter{equation}{0}
\setcounter{figure}{0}
\setcounter{section}{0}

\makeatletter
\renewcommand \thesection{S\@arabic\c@section}
\renewcommand{\theequation}{S\arabic{equation}}
\renewcommand{\thefigure}{S\arabic{figure}}

\textbf{Exact solution of the unitary case --}
We derive explicitly the probability distribution $P_t(x|x_0)$ of an arc endpoint under unitary Clifford Gaussian evolution. Then, we use it to evaluate the growth of the entanglement entropy.

We are interested in determining the distribution $P_t(x) = P_t(x|x_0)$, where we omit the starting conditions for brevity. First, we observe that for any $t>0$ we have $P_t(2n-1) = P_t(2n)$ for any physical site $n$, and thus we introduce $\mathcal{P}_t(n) = P_t(2n-1) + P_t(2n)$. This comes from the random permutations implemented by gates, which immediately make positions $2n$ and $2n+1$ equally likely as soon as the first unitary is applied.

Let us assume that $t=t_\mathrm{o}$ is odd, such that the last unitary layer applied couples each even physical site with its right neighbour. If $n$ is even, the Majorana at $x$ at time $t$ must have come from either the physical site $n$ or from $n+1$ at $t-1$; similarly, if $n$ is odd then the physical site at the previous step must have been either $n-1$ or $n$. Since the gate implements a random permutation, all Majorana paths are equally likely and thus $\mathcal{P}_{t_\mathrm{o}}(n)$ is just the average of the probabilities of the two possible previous sites at $t_\mathrm{o}-1$. This gives us the recursion
\begin{equation}
    \mathcal{P}_{t_\mathrm{o}}(n) = \frac{1}{2}\bigg[\mathcal{P}_{t_\mathrm{o}-1}(2\lfloor n/2 \rfloor) + \mathcal{P}_{t_\mathrm{o}-1}(2\lfloor n/2 \rfloor+1)\bigg].
\end{equation}
Analogously, for even $t=t_\mathrm{e}$ we have
\begin{equation}
    \mathcal{P}_{t_\mathrm{e}}(n) = \frac{1}{2}[\mathcal{P}_{t_\mathrm{e}-1}(2\lfloor (n+1)/2 \rfloor-1) + \mathcal{P}_{t_\mathrm{e}-1}(2\lfloor (n+1)/2 \rfloor)].
\end{equation}

These equations give rise to a binomial probability distribution that spreads over time. In detail, we obtain
\begin{equation}\label{unitary_solution}
\begin{split}
    P_t(n)
    = \begin{cases}
        \frac{1}{2^{t+1}} b(t-1,(t-1)/2+\lfloor (x-1)/4 \rfloor - \lfloor (x_0-1)/4 \rfloor) \quad\text{for odd}\, $t$,\\
        \frac{1}{2^{t+1}} b(t-1,t/2-1+\lfloor (x+1)/4 \rfloor - \lfloor (x_0-1)/4 \rfloor) \quad\text{for even}\, $t$,
    \end{cases}
    \end{split}
\end{equation}
where $b(n,k)$ is the binomial coefficient $\binom{n}{k}$ for $0\leq k \leq n$ and it is zero otherwise. It can be checked by explicit substitution that Eq.~\eqref{unitary_solution} satisfies the recursion relations with the correct initial condition. At long times, this distribution is well approximated by a Gaussian centered around $x=x_0$ with variance $4t$, i.e., $P_t(x|x_0) \approx \mathcal{N}(x;x_0,4t)$.

As mentioned in the main text, the entanglement entropy is given by half the number of arcs that connect the two subsystems. For a half-chain bipartition, we need to count the average number of arcs with an endpoint in $[1,N]$ and the other in $[N+1,2N]$. In the following, we consider the limit of $N\gg \sqrt{t}\gg 1$, in which case it is reasonable to assume that the two endpoints of an arc, starting from $x_0=2n$ and $x_0=2n+1$ at $t=0$, are independently distributed at $t$. This yields
\begin{equation}
    \overline{S_{N/2}(t)} = \frac{1}{2}\sum_{n=1}^{N}\bigg[\sum_{x=1}^{N} P_t(x|2n-1)\sum_{y=N+1}^{2N}P_t(y|2n)
    +\sum_{x=N+1}^{2N} P_t(x|2n-1)\sum_{y=1}^{N}P_t(y|2n) \bigg].
\end{equation}
Taking the continuum limit and using the Gaussian approximation of $P_t(x|x_0)$, after some straightforward passages we find
\begin{equation}
    \overline{S_{N/2}(t)} = \sqrt{\frac{t}{2}}\int_0^A dx_0 [\mathrm{erf}(A+x_0)-\mathrm{erf}(x_0)]
    [\mathrm{erf}(A-x_0)+\mathrm{erf}(x_0)],
\end{equation}
where $A = N/\sqrt{8t} \gg 1$. The final integral can be evaluated explicitly for $A\to \infty$ using $\mathrm{erf}(A+x_0)\approx 1$ and $\mathrm{erf}(A-x_0)\approx 1$ near $x_0=0$. The result is $\overline{S_{N/2}(t)} = \sqrt{t/\pi}$ at leading order, which matches well with our numerics.

\textbf{Clifford Gaussian Page Curve --}
Here we derive an explicit formula for the bipartite entanglement entropy $\overline{S_i}$, averaged over the set of Gaussian stabilizer states. Thanks to the mapping to the arc model, sampling a random Gaussian stabilizer state is equivalent to selecting a random configuration of the $N$ arcs. As a straightforward combinatorial exercise, one finds that the total number of such arc configurations is given by $a_N = \frac{(2N)!}{N! \, 2^N}$. The problem now reduces to counting how many of these configurations contain a given number of arcs crossing between the region $A = \{1, 2, \dots, 2i\}$ and its complement $\overline{A} = \{2i+1, 2i+2, \dots, 2N\}$. This number must be even, so we denote it by $2j$ (if the number was odd, it would be impossible to pair all the sites in $A$ that are not connected to $\overline{A}$ among themselves). There are respectively $\binom{2i}{2j}$ and $\binom{2N-2i}{2j}$ ways of selecting the points to connect in $A$ and in $\overline{A}$ respectively, and $(2j)!$ different ways of connecting them once selected. The unpaired sites in $A$, $\overline{A}$ can be arranged in $a_{i-j}$ and $a_{N-i-j}$ different configurations, respectively. Overall, this yields 
\begin{equation}
      [\text{\# configurations with $2j$ arcs connecting $A$ and $\overline{A}$}] =\binom{2i}{2j} \, \binom{2N-2i}{2j} \, (2j)! \, a_{i-j} \, a_{N-i-j} \, . 
\end{equation}
The average entanglement is therefore obtained by summing over all possible values of $2j$ in the range $[0, \min(2N,2N-2i)]$, which gives
\begin{equation}
    \overline{S_i} =\frac{1}{2 \, a_N} \sum_{j=0}^{\min(i,N-i)} \bigg[ 2j \binom{2i}{2j} \binom{2N-2i}{2j} (2j)! a_{i-j} a_{N-i-j} \bigg] \, .
\end{equation}
Finally, expanding this equation for large $N$ fixing $i=N/2$ yields $\overline{S_{N/2}}=\frac{N}{4} +\frac{1}{8}+O(N^{-1})$.

\textbf{Master equation for the monitored Clifford Gaussian dynamics --}
Here we derive the master equation for the arc-length distribution $P_t(\ell)$ introduced in the main text, and we proceed to solve it. Given an arc with endpoints $(\mu,\nu)$, let $x=\lfloor{(\mu+1)/2\rfloor}$ and $y=\lfloor{(\nu+1)/2\rfloor}$ the corresponding physical lattice sites. We are interested in their joint probability distribution. In presence of measurements, each timestep of the dynamics contains a unitary layer and a measurement layer. Let $P_t(x,y)$ and $P_{t+1/2}(x,y)$ be the endpoint distributions after a unitary or measurement layer, respectively. At long times, it is reasonable to assume that arcs delocalize, losing dependence on their center of mass position. As a consequence, we assume $P_t(x,y) = P_t(y-x)$ and $P_{t+1/2}(x,y)=P_t(y-x)$, where $y-x=\ell$ is the arc length, taken with the sign. 

Let us consider the evolution of the probability distribution under a unitary step. As discussed in the unitary case, both $x$ and $y$ either remain unchanged or shift by $\pm 1$. On average, each endpoint will cause the arc length to change by $\pm 1$ with probability $1/4$ and to be unaffected otherwise. This yields the update rule $P_{t}(\ell) = \left[f * (f * P_{t-1/2})\right](\ell)$, where $f(x) = (\delta_{x,-1}+2\delta_{x,0}+\delta_{x,1})/4$, and $*$ denotes a convolution. Notice that the convolution is applied twice, once for each arc endpoint.

Next, consider the measurement layer. Here we assume a small measurement density $p\ll 1$, allowing us to safely neglect all processes in which an arc is affected by more than one measurement. As explained in the main text, a measurement glues together two arcs of length $\ell$ and $\ell'$ into a composite arc of length $\ell+\ell'$, and creates a new localized arc of vanishing length (because its Majorana endpoints belong to the same physical site). This happens with a probability $1-(1-p)^2\approx 2p$ (because there are $2$ endpoints that can be measured). Since we are keeping track of a single-arc distribution and this process involves two arcs simultaneously, we model a measurement as follows: half of the time, the original arc is mapped to the composite arc of length $\ell+\ell'$, whereas the other half it is mapped to the local arc of zero length. Overall, this yields the following possibilities. (i) No measurement: with probability $1-2p$, no measurement occurs, and an arc of length $\ell$ remains unchanged. (ii) Measurement, composite arc: with probability $2p\cdot 1/2$, the arc of length $\ell$ is originated by the composition of two arcs of lengths $\ell_1$ and $\ell_2$ such that $\ell = \ell_1+\ell_2$. These sizes are picked randomly from their current distributions, i.e.\ $P_{t}(\ell_{1})$,$P_{t}(\ell_{2})$. (iii) Measurement, localized arc: with probability $2p\cdot 1/2$, $\ell=0$ is the local arc generated by a measurement.

Overall, we obtain $P_{t+1/2}(\ell) = (1-2p)P_{t}(\ell) + p [\delta_{\ell,0} + \sum_{\ell_1,\ell_2}P_{t}(\ell_1)P_{t}(\ell_2)\delta_{\ell,\ell_1+\ell_2}]$. 
By combining the update equations for the probability distribution, we find
\begin{equation}
    P_{t+1}(\ell) = (1-2p)\left[f*(f*P_t)\right](\ell) + p(f * f)(\ell) + p\left[f*\left[f*\left(P_t*P_t\right)\right]\right](\ell) \equiv \mathcal{K}[P_t](\ell),
\end{equation}
which is a non-linear discrete master equation. Given the convolutions, it simplifies dramatically in momentum space (assuming translational symmetry), so we introduce $\Tilde{P}_t(k) = \sum_\ell e^{-i k \ell} P_t(\ell)$ and $\Tilde{f}(k) = \sum_\ell e^{-i k \ell} f(\ell) = \cos^2(k/2)$, yielding $\Tilde{P}_{t+1}(k) = \cos^4(k/2)[(1-2p)\Tilde{P}_t(k) + p + p \Tilde{P}_t^2(k)]$. 
If we are interested in the steady state, we assume $\Tilde{P}_{t}(k) = \Tilde{P}_{t+1}(k) \equiv \Tilde{P}(k)$, which provides directly Eq.~\eqref{master_equation_k} of the main text.
This is a quadratic equation that can be solved straightforwardly, giving
\begin{equation}\label{ME_solution}
    \Tilde{P}(k) = -\frac{1-2p-\cos^{-4}(k/2)}{2p} - \frac{\sqrt{[1-\cos^{-4}(k/2)][1-4p-\cos^{-4}(k/2)]}}{2p}.
\end{equation}
The second solution shows $\Tilde{P}(k)>1$, and is thus unphysical. Most importantly, the behavior of Eq.~\eqref{ME_solution} for $|k|\ll 1$ determines the asymptotic decay of $P(\ell)$ in real space. After expanding, we find $\Tilde{P}(k)\approx 1-|k|\sqrt{2p}$. The inverse Fourier transform finally gives the asymptotic decay $P(\ell) = \int_{-\pi}^\pi \frac{dk}{2\pi} e^{i k \ell} \Tilde{P}(k) \xrightarrow{\ell\gg 1} \frac{1}{\pi\sqrt{2p}\ell^2}$, 
which is obtained by explicitly evaluating the contribution of $|k|\ll 1$ to the integral.

Finally, the average half-chain entanglement entropy is related to the arc-length distribution by
\begin{equation}
    \overline{S_{N/2}} = \frac{1}{2}\sum_{x=1}^{N/2}\sum_{y=N/2+1}^{2N}\left[P(x,y) +P(y,x)\right].
\end{equation}
After replacing the sums by integrals and inserting $P(\ell)\approx1/\ell^2$, we obtain the logarithmic scaling $\overline{S_{N/2}}\sim\log N$.

\end{document}